\documentclass{kapproc} 

\usepackage{procps} 

\usepackage[dvips]{graphicx}

\upperandlowercase

\kluwerbib 

\begin{document}

\articletitle[Dusty Starbursts and the Growth of Cosmic Structure]
{Dusty Starbursts and the Growth of Cosmic Structure}

\author{D. Elbaz, E. Moy}

\affil{CEA Saclay$/$DSM$/$DAPNIA$/$Service d'Astrophysique\\
Orme des Merisiers, F-91191 Gif-sur-Yvette Cedex\\
France}
\email{delbaz@cea.fr}


\anxx{Elbaz\, David}

\begin{abstract}
Dusty starbursts were more numerous around $z\sim$ 1 than today and
appear to be responsible for the majority of cosmic star formation
over the Hubble time.  We suggest that they represent a common phase
within galaxies in general which is triggered by the growth of cosmic
structure.
\end{abstract}


\section{Introduction}
Dusty starbursts producing stars at a rate of about 50 M$_{\odot}$
yr$^{-1}$ were very efficiently detected by ISOCAM onboard the
Infrared Space Observatory (ISO) at 15\,$\mu$m below $z\sim$ 1.3 where
the broad bump due to aromatic features in the mid-IR (PAHs,
polycyclic aromatic features) remains within the ISOCAM filter. These
luminous infrared (LIR) galaxies with infrared luminosities greater
than $L_{\rm IR}(8-1000\,\mu {\rm m})=\,10^{11}$ L$_{\odot}$ were
producing more than thirty times their present-day comoving infrared
luminosity density at $z\sim$ 1 than today, when it is only a few
percent of the bolometric luminosity radiated by galaxies in the
optical to near-infrared (Elbaz {\it et al.} 2002, Chary \& Elbaz
2001). More distant LIR galaxies were detected in the sub-millimeter
with SCUBA (Chapman et al. 2003 and references therein), with IR
luminosities of several 10$^{12}$ L$_{\odot}$ due to sensitivity
limits, but their large number density confirms the same trend as
observed with ISO, i.e. that the bulk of the cosmic history of star
formation was dominated by dusty star formation and can only be
derived when carefully accounting for this dusty starbursting phase
taking place in most if not all galaxies. This is also suggested by
the fact that these galaxies contribute to about two thirds of the
cosmic infrared background (CIRB, Elbaz et al. 2002), which contains
about half or more of the total energy radiated by galaxies through
the Hubble time, and dominate the cosmic history of star formation
(Chary \& Elbaz 2001).  Are these galaxies dominated by star formation
or nuclear activity ? What is triggering their strong activity ? Is it
triggered by external interactions or did it happen naturally within
isolated galaxies ?

\begin{figure} 
\includegraphics[width=12cm]{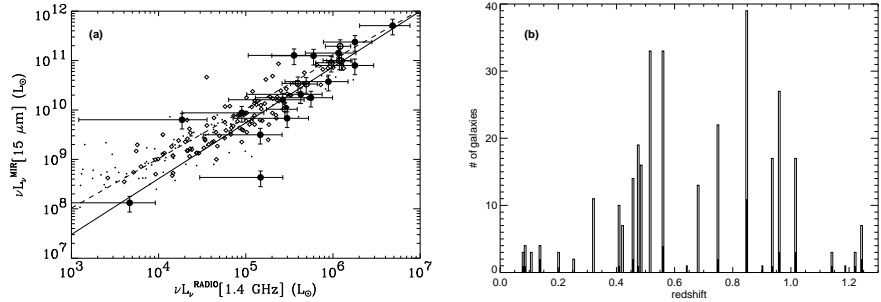}
\caption{{\bf a)} 15\,$\mu$m versus radio continuum (1.4 GHz)
rest-frame luminosities. Small filled dots: sample of 109 local
galaxies from ISOCAM and NVSS. Filled dots with error bars: 17 HDFN
galaxies ($z\sim$0.7, radio from VLA or WSRT). Open dots with error
bars: 7 CFRS-14 galaxies ($z\sim$0.7, Flores et al. 1999, radio from
VLA). Open diamonds: 137 ELAIS galaxies ($z\sim$ 0-0.4). {\bf a)}
histogram of the redshift distribution of field (white) and ISOCAM
(dark) galaxies which belong to redshift peaks as defined by Cohen et
al. (2000).}
\label{FIG:radio}
\end{figure}

\section[]{Origin of the luminosity of luminous infrared galaxies at z=1}
Among the fields surveyed in the mid-IR with ISOCAM, the Hubble Deep
Field North (HDFN) and its Flanking Fields (total area of 27
$arcmin^2$), is the best one to study the contribution of active
galactic nuclei (AGN) to the IR luminosity of these LIR
galaxies. Thanks to the deepest soft to hard X-ray survey ever
performed with Chandra in the HDFN, it is possible to pinpoint AGNs
including those affected by dust extinction. Among a total number of
95 sources detected at 15\,$\mu$m in this field (Aussel et al. 1999,
2003 in prep.; with 47 above a completeness limit of 0.1 mJy), only 5
sources were classified as AGN dominated on the basis of their X-ray
properties (Fadda et al. 2002). Hence, unless a large number of AGNs
are so dust obscured that they were even missed with the 2 Megaseconds
Chandra survey, the vast majority of ISOCAM LIR galaxies are powered
by star formation. We are then left with the following question: how
can one derive a total IR luminosity on the sole basis of a mid-IR
measurement ? We showed in Elbaz et al. (2002) that in the local
universe, there exists a very strong correlation between the mid-IR
and total IR luminosity of galaxies over three decades in
luminosity. In order to test whether these correlations remain valid
up to $z\sim$ 1, it is possible to use another correlation existing
between the radio and total IR luminosity in the local universe and
check whether both the mid-IR and radio give a consistent prediction
for the total IR luminosity. In the Fig.~\ref{FIG:radio}a, we have
reproduced the plot from Elbaz et al. (2002) complemented with
galaxies detected within the ELAIS survey (Rowan-Robinson et
al. 2003). Except at low luminosities were the contribution of cirrus
to the IR luminosity becomes non negligible, the 1.4 GHz and
15\,$\mu$m rest-frame luminosities are correlated up to $z\sim$ 1 and
therefore predict very consistent total IR luminosities from which
star formation rates as well as the contribution of these objects to
the CIRB can be computed, leading to the results mentioned in the
introduction.

\begin{figure}[t]
\includegraphics[width=12cm]{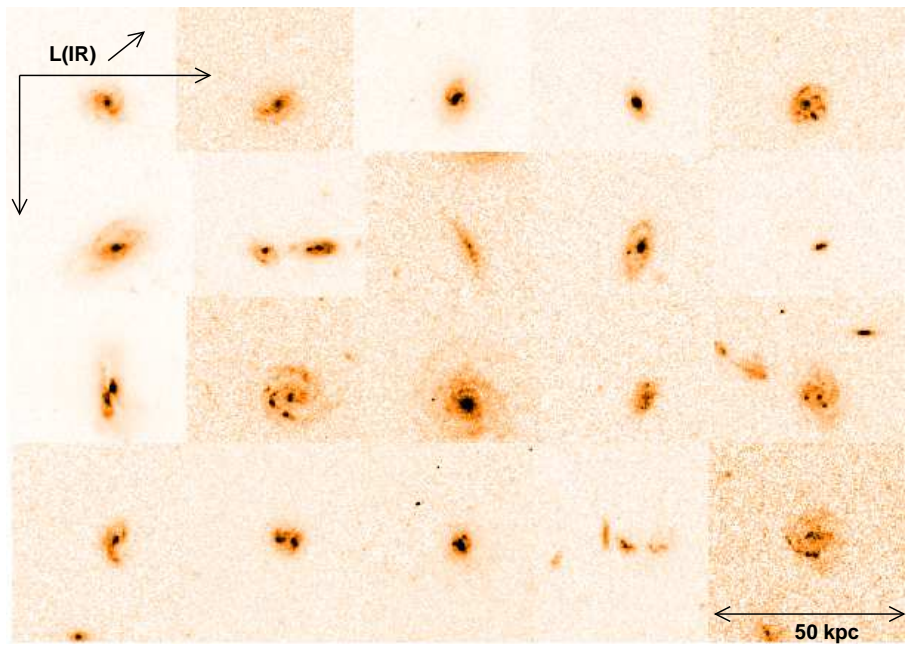}
\caption{HST-ACS images of LIR galaxies with 11 $\leq$ log(L$_{\rm
IR}/L_{\odot}$) $\leq$ 12 (LIRGs) and $z\sim$ 0.7. The double-headed
arrow indicates the physical size of 50 kpc. The IR luminosity
increases from left to right and from top to bottom.}
\label{FIG:camgal}
\end{figure}

The next question concerns the physical origin of this dusty starburst
phase in galaxies (see Elbaz \& Cesarsky 2003). In order to address it
we have plotted in the Fig~\ref{FIG:radio}b the histogram of the
redshift distribution of the ISOCAM galaxies with a spectroscopic
redshift and above the completeness limit of 0.1 mJy in the HDFN. For
comparison, only the redshifts where Cohen et al. (2000) found a
redshift peak is plotted for field galaxies.  One can clearly see that
except for 3 ISOCAM galaxies, all are located within these redshift
peaks. We have performed Monte-Carlo simulations to compute the
probability that a random sample of optical galaxies of the same size
than that of ISOCAM galaxies would fall in as many redshift peaks and
we found that it is of the order of one percent, and even less if we
restrict ourselves to the biggest redshift peaks (Moy et al., in
prep). Is this result in agreement with expectations ?  As showed by
Franceschini et al. (2001), ISOCAM galaxies are relatively massive,
hence more subject to belong to denser regions. Moreover LIR galaxies
are mostly triggered by interactions in the local universe and one
would naturally expect the same to take place in the more distant
universe, hence in connection with cosmic structure formation. Thanks
to the deepest existing survey with the ACS camera onboard the HST
(GOODS survey, Giavalisco et al. 2003), we were able to study the
morphology of ISOCAM galaxies in this field centered on the HDFN. A
first result follows our expectations: most of these galaxies present
a disturbed morphology and this is truer as a function of increasing
IR luminosity (see Fig.~\ref{FIG:camgal}). A second result was less
expected: several LIR galaxies appear to be relatively big disk
galaxies, which could be affected by a neighboring dwarf galaxy or a
passing-by galaxy but they are not major mergers in the main phase of
interacting. This is a very important result that should be carefully
considered by authors of galaxy formation simulations. A large
fraction of present-day stars may originate from a phase of violent
star formation triggered by passing-by galaxies even within disk
galaxies which could remain as such in the present-day
universe. Another important consequence of this type of physical
origin for the dusty starburst phase is that there are not enough
major mergers in simulations to explain the large excess of LIR
galaxies in the distant universe and the fact that they dominate the
cosmic star formation history. Indeed each individual galaxy could
have experienced several encounters with other galaxies at the time of
structure formation which would induce a series of bursts of star
formation, each of them responsible for only a few percent of the
final stellar mass of the galaxy.

\begin{chapthebibliography}{1}

Aussel, H., Cesarsky, C.J., Elbaz, D., Starck, J.L. 1999, A\&A 342, 313

Chapman, S.C., Blain, A.W., Ivison, R.J., Smail, I. 2003, Nature 422, 695

Chary, R.R., Elbaz, D. 2001, ApJ 556, 562

Cohen, J.G., Hogg, D.W., Blandford, R., et al. 2000, ApJ 538, 29 

Elbaz, D., Cesarsky, C.J., Fadda, D., et al. 1999, A\&A 351, L37

Elbaz, D., Cesarsky, C.J., Chanial, P., et al. 2002, A\&A, 384, 848

Elbaz, D., Cesarsky, C.J. 2003, Science 300, 270

Fadda, D., Flores, H., Hasinger, G., et al. 2002, A\&A 383, 838

Flores, H.,  Hammer, F., Thuan, T.X., et al. 1999, ApJ 517, 148

Franceschini, A., Aussel, H., Cesarsky, C.J., Elbaz, D., Fadda, D. 2001, A\&A 378, 1

Giavalisco, M.,  et al. 2003, ApJ (in press, astro-ph$/$0309105)

Rowan-Robinson, M., et al. 2003, MNRAS (submitted, astro-ph$/$0308283)

\end{chapthebibliography}

\end{document}